\documentclass{jfm}

\newcommand{\EQ}{\begin{equation}}
\newcommand{\EN}{\end{equation}}
\newcommand{\EQA}{\begin{eqnarray}}
\newcommand{\ENA}{\end{eqnarray}}

\newcommand{\Sec}[1]{section~\ref{#1}}
\newcommand{\Secs}[2]{sections~\ref{#1} and \ref{#2}}

\newcommand{\fig}[1]{figure~\ref{#1}}
\newcommand{\figs}[2]{figures~\ref{#1} and \ref{#2}}

\newcommand{\tab}[1]{table~\ref{#1}}

\newcommand{\mean}[1]{\overline #1}

{}
{}
{}

\newcommand{\meanuu}{\overline{\mbox{\boldmath $u$}}{}}{}
{}
{}
{}
{}
{}
{}
{}
{}
{}
{}
{}
{}
{}
{}
{}
{}

{}

{}

{}
\newcommand{\Rgas}{{\cal R}}{}
{}

\newcommand{\eee}{\hat{\mbox{\boldmath $e$}} {}}

\newcommand{\uuu}{{\bm u}}

\newcommand{\FFF}{\mbox{\boldmath ${\cal F}$} {}}

\def\Ma{\mbox{\rm Ma}}

\def\Pra{\mbox{\rm Pr}}

\def\Rey{\mbox{\rm Re}}

\def\Pe{\mbox{\rm Pe}}

\def\cP{c_{P}}
\def\cV{c_{V}}

\def\cs{c_{\rm s}}

\def\kf{k_{\rm f}}

\def\urms{u_{\rm rms}}

\def\nut{\nu_{\rm t}}

\def\chit{\chi_{\rm t}}
\def\chitz{\chi_{\rm t0}}

\def\nutz{\nu_{\rm t0}}

\def\onethird{{\textstyle{1\over3}}}

\newcommand{\Fenthz}{{F^{\rm enth}_z}}
\newcommand{\Ryz}{{R_{yz}}}
\newcommand{\pd}{\partial}
\newcommand{\SSt}{\bm{\mathsf{S}}}
\newcommand{\SStij}{\mathsf{S}_{ij}}
\def\onehalf{{\textstyle{1\over2}}}
\newcommand{\calR}{{\cal R}}
\newcommand{\brac}[1]{\langle #1 \rangle}

\usepackage{graphicx}
\usepackage{newtxtext}
\usepackage{newtxmath}
\usepackage{natbib,bm}
\usepackage{hyperref}
\usepackage{comment}
\hypersetup{
    colorlinks = true,
    urlcolor   = blue,
    citecolor  = blue,
}
\graphicspath{{./fig/}{./png/}}

\newcommand{\RomanNumeralCaps}[1]

\title{Turbulent Prandtl number from isotropically forced turbulence}

\author{Petri J. K\"apyl\"a\aff{1}
  \corresp{\email{pkaepyl@uni-goettingen.de}}, \and Nishant K. Singh\aff{2}}

\affiliation{
  \aff{1} Institute for Astrophysics and Geophysics,
  G\"ottingen University, Friedrich-Hund-Platz 1, 37077 G\"ottingen, Germany
  \aff{2} Inter-University Centre for Astronomy and Astrophysics, Post Bag 4
  Ganeshkhind, Savitribai Phule Pune University Campus, Pune 411 007,
  India}

\begin{document}
\maketitle

\begin{abstract}
Turbulent motions enhance the diffusion of large-scale flows and
temperature gradients. Such diffusion is often parameterized by
coefficients of turbulent viscosity ($\nut$) and turbulent thermal
diffusivity ($\chit$) that are analogous to their microscopic
counterparts. We compute the turbulent diffusion coefficients by
imposing large-scale velocity and temperature gradients on a turbulent
flow and measuring the response of the system. We also confirm our
results using experiments where the imposed gradients are allowed to
decay. To achieve this, we use weakly compressible three-dimensional
hydrodynamic simulations of isotropically forced homogeneous
turbulence. We find that the turbulent viscosity and thermal
diffusion, as well as their ratio the turbulent Prandtl number,
$\Pra_{\rm t} = \nut/\chit$, approach asymptotic values at
sufficiently high Reynolds and Pecl\'et numbers. We also do not find a
significant dependence of $\Pra_{\rm t}$ on the microscopic Prandtl
number $\Pra = \nu/\chi$. These findings are in stark contrast to
results from the $k-\epsilon$ model which suggests that $\Pra_{\rm t}$
increases monotonically with decreasing $\Pra$. The current results
are relevant for the ongoing debate of, for example, the nature of the
turbulent flows in the very low $\Pra$ regimes of stellar convection
zones.
\end{abstract}



\section{Introduction}
\label{sec:intro}

The fluids in stellar convection zones are generally characterized by
a low microscopic Prandtl number, $\Pra = \nu/\chi$, where $\nu$ is
the kinematic viscosity and $\chi$ is the thermal diffusivity
\citep[e.g.][]{O03,2019ApJ...876...83A}. Typical values in the bulk of
the solar convection zone, for example, range between $10^{-6}$ and
$10^{-3}$ \citep[][]{2020RvMP...92d1001S}. Recently, several studies
have explored the possibility that solar convection operates at a high
effective Prandtl number regime, meaning that the turbulent Prandtl
number $\Pra_{\rm t}$ exceeds unity
\citep[e.g.][]{2016AdSpR..58.1475O,2017ApJ...851...74B,2018PhFl...30d6602K},
as a possible solution to the too high velocity amplitudes in
simulations in comparison to the Sun
\citep[e.g.][]{HDS12,2020RvMP...92d1001S}. However, few attempts have
been made to actually measure the turbulent Prandtl number from
simulations. A notable exception is the study of
\cite{2021PhRvF...6j0503P} who reported that the turbulent Prandtl
number decreases steeply as a function of the molecular Prandtl number
such that $\Pra_{\rm t} \propto \Pra^{-1}$ in simulations of standard
Boussinesq and variable heat conductivity Boussinesq convection.

Here we set out to measure turbulent viscosity and thermal diffusivity
from a simpler system of isotropically forced homogeneous
turbulence. This is done by imposing large-scale gradients of velocity
and temperature (equivalently specific entropy) and measuring the
response of the system. The turbulent diffusion coefficients are
computed from the Boussinesq ansatz and an analogous expression for
the enthalpy flux. This method provides a direct measurement of the
diffusion coefficients without the need to resort to turbulent
closures. Similar methods were used recently to measure the turbulent
magnetic Prandtl number \citep{2020A&A...636A..93K}. We compare our
results with those from the widely used $k-\epsilon$ model which was
also used by \cite{2021PhRvF...6j0503P}. We show that the direct
results and those from the $k-\epsilon$ model are systematically
different and that the latter yields misleading results.

\section{The model}
\label{sec:model}

We model isotropically forced, non-isothermal, turbulence in a fully
periodic cube of volume $(2\pi)^3$. We solve the equations of fully
compressible hydrodynamics
\begin{eqnarray}
\frac{D \ln \rho}{D t} &=& -\bm\nabla \bm\cdot \uuu, \label{equ:dens}\\
\frac{D\uuu}{D t} &=& -\frac{1}{\rho}(\bm\nabla p - \bm\nabla\bm\cdot 2 \nu \rho \bm{\mathsf{S}}) + {\bm f} - \frac{1}{\tau}(\uuu - \meanuu_0) ,\label{equ:mom} \\
T \frac{D s}{D t} &=& -\frac{1}{\rho}\left(\bm\nabla\bm\cdot\FFF_{\rm rad}-{\cal C} \right) + 2 \nu \bm{\mathsf{S}}^2 - \frac{T}{\tau}(s - \mean{s}_0),
\label{equ:ent}
\end{eqnarray}
where $D/Dt = \pd/\pd t + \uuu\cdot\bm\nabla$ is the advective
derivative, $\rho$ is the density, $\uuu$ is the velocity, $p$ is the
pressure, $\nu$ is the kinematic viscosity, $\SSt$ is the traceless
rate-of-strain tensor with
\begin{eqnarray}
\SStij = \onehalf (u_{i,j} + u_{j,i}) - \onethird \delta_{ij} \bm\nabla\bm\cdot\uuu,\label{equ:SSt}
\end{eqnarray}
${\bm f}$ is the external forcing, $\tau$ is a relaxation timescale,
and $\meanuu_0$ is the target mean velocity profile. Furthermore, $T$
is the temperature, $s$ is the specific entropy, $\FFF_{\rm rad}$ is
the radiative flux, ${\cal C}$ is a cooling term, and $\mean{s}_0$ is
the target mean specific entropy profile. Radiation is
modeled via the diffusion approximation, with the radiative flux
given by
\begin{eqnarray}
\FFF_{\rm rad} = -\cP \rho \chi \bm\nabla T,
\label{equ:Frad}
\end{eqnarray}
where $\cP$ is the specific heat in constant pressure and $\chi$
is the thermal diffusivity. The ideal gas equation of state $p=(\cP -
\cV) \rho T =\calR \rho T$ is assumed, where $\calR$ is the gas
constant, and $\cV$ is the specific heat capacity at constant
volume. In the presence of an imposed large-scale flow, viscous
dissipation of kinetic energy acts as a source for
thermal energy and leads to a linear increase of the
temperature. Additional volumetric cooling is applied to counter this
with
\begin{eqnarray}
{\cal C}({\bm x}) = \rho \cP \frac{T({\bm x}) - \brac{T_0}}{\tau_{\rm
    cool}},
\label{equ:cool}
\end{eqnarray}
where $\brac{T_0}$ is the volume-averaged initial temperature and
$\tau_{\rm cool}$ is a cooling timescale. We use $\tau = \tau_{\rm
  cool} = (c_{s0}k_1)^{-1}$, where $c_{s0}$ is the initial uniform
value of the sound speed and $k_1$ is the wavenumber corresponding to
the box scale.

The external forcing is given by \citep[see][]{B01}
\begin{eqnarray}
  {\bm f} = \Real\{ N(t) {\bm f}_{{\bm k}(t)} \exp[{\rm i} {\bm k}(t)\bm\cdot {\bm x} - {\rm i}\phi(t)]\},
\end{eqnarray}
where ${\bm x}$ is the position vector and $N(t)= f_0 \cs
(k(t)\cs/\delta t)^{1/2}$ is a normalization factor where $f_0$ is the
forcing amplitude, $k=|{\bm k}|$, $\delta t$ is the length of the time
step, and $-\pi < \phi(t) < \pi$ is a random delta-correlated
phase. The vector ${\bm f}_{\bm k}$ describes non-helical transverse
waves, and is given by
\begin{eqnarray}
{\bm f}_{\bm k} = \frac{{\bm k} \times \hat{\bm e}}{\sqrt{ {\bm k}^2 - ({\bm k} \bm\cdot \hat{\bm e})^2 }},
\end{eqnarray}
where $\hat{\bm e}$ is an arbitrary unit vector, and where the
wavenumber ${\bm k}$ is randomly chosen. The target profiles of mean
velocity and specific entropy are given by
\begin{eqnarray}
  \meanuu_0 &=& u_0\sin(k_1 z){\eee}_y, \label{equ:impu} \\
  \mean{s}_0 &=& s_0 \sin(k_1 z). \label{equ:imps}
\end{eqnarray}
In addition to the physical diffusion, the advective terms in
(\ref{equ:dens}) to (\ref{equ:ent}) are implemented in terms of
fifth-order upwinding derivatives with sixth-order hyperdiffusive
corrections and flow-dependent diffusion coefficients; see Appendix~B
of \cite{DSB06}.

The {\sc Pencil Code}
\citep{2021JOSS....6.2807P}\footnote{\href{http://github.com/pencil-code}{http://github.com/pencil-code}},
which uses high-order finite differences for spatial and temporal
discretisation, was used to produce the numerical simulations.

\subsection{Units, system parameters, and diagnostics}

The equations are non-dimensionalized by choosing the units
\begin{eqnarray}
[x] = k_1^{-1},\ \ [\rho] = \rho_0,\ \ [u] = c_{\rm s0},\ \ [s] = \cP,
\end{eqnarray}
where $\rho_0$ is the initial uniform density and $c_{\rm s0} =
\sqrt{\gamma \Rgas T_0}$ is the sound speed corresponding to the initial
temperature $T_0$.
The level of velocity fluctuations is determined by the forcing
amplitude $f_0$ along with the kinematic viscosity. A key system
parameter is the ratio of kinematic viscosity and thermal diffusion
or the Prandtl number
\begin{eqnarray}
\Pra = \frac{\nu}{\chi},
\end{eqnarray}
which is varied between $0.01$ and $10$ in the present study. The
Reynolds and P\'eclet numbers quantify the level of turbulence of the
flows:
\begin{eqnarray}
\Rey = \frac{\urms}{\nu \kf},\ \ \ \Pe = \Pra\Rey = \frac{\urms}{\chi \kf},
\end{eqnarray}
where $\urms = \sqrt{\brac{(\uuu - \mean{\uuu}_0)^2}}$ is the
volume-averaged fluctuating rms-velocity and $\kf$ is the average
forcing wavenumber characterizing the energy injection scale.  The
latter is chosen from a uniformly distributed narrow range in the
vicinity of $5k_1$. The imposed gradients of large-scale flow and
entropy are quantified by
\begin{eqnarray}
\Ma_s = \frac{u_0}{c_{\rm s0}}, \ \ \Ma_g = \frac{[(\gamma-1)s_0 T_0]^{1/2}}{c_{\rm s0}},
\end{eqnarray}
where $\Ma_s$ is the Mach number of the mean flow and $k_U = k_s =
k_1$. The Mach number of the turbulent flow is given by
\begin{eqnarray}
\Ma = \frac{\urms}{\brac{\cs}},
\end{eqnarray}
where $\brac{\cs}$ is the volume-averaged speed of sound.

\noindent
Mean values are taken to be horizontal averages denoted by overbars,
that is
\begin{eqnarray}
  \mean{f} = \frac{1}{(2\pi)^2} \int_x \int_y f({\bm x}) dx dy.
\end{eqnarray}
Often an additional time average over the statistically steady part of
the simulation is taken. Volume averages are denoted by angle brackets
$\brac{.}$ apart from the rms-values which are always assumed to be
volume-averaged unless otherwise stated. Errors were estimated by
dividing the time series in three parts and averaging over each
subinterval. The greatest deviation from the average computed
over the whole time series was taken as the error estimate.

\begin{table}
\setlength{\tabcolsep}{7pt}
\scriptsize
  \begin{center}
\def~{\hphantom{0}}
  \begin{tabular}{cccccccc}
Set & $\Pra$ & $\Rey$ & $\Pe$ & $\Ma$ & $\Ma_s$ & $\Ma_g$ & \# runs \\
\hline
\mbox{i001} & $0.01$ & $27 \ldots 779$  & $0.27 \ldots 7.8$  & $0.069 \ldots 0.079$ & $0.01 \ldots 0.03$ & $0.07 \ldots 0.1$ & $10$ \\
\mbox{i002} & $0.02$ & $27 \ldots 779$  & $0.54 \ldots 16$   & $0.069 \ldots 0.080$ & $0.01 \ldots 0.03$ & $0.07 \ldots 0.1$ & $12$ \\
\mbox{i005} & $0.05$ & $27 \ldots 780$  & $1.4  \ldots 39$   & $0.069 \ldots 0.080$ & $0.01 \ldots 0.03$ & $0.07 \ldots 0.1$ & $12$ \\
\mbox{i010} & $0.1$  & $12 \ldots 800$  & $1.2  \ldots 80$   & $0.060 \ldots 0.082$ & $0.01 \ldots 0.03$ & $0.07 \ldots 0.1$ & $30$ \\
\mbox{i020} & $0.2$  & $27 \ldots 781$  & $5.4  \ldots 156$  & $0.069 \ldots 0.080$ & $0.01 \ldots 0.03$ & $0.07 \ldots 0.1$ & $10$ \\
\mbox{i025} & $0.25$  & $397$  & $99$  & $0.081$ & $0.01 \ldots 0.03$ & $0.07 \ldots 0.1$ & $2$ \\
\mbox{i050} & $0.5$  & $27 \ldots 781$  & $14   \ldots 390$  & $0.069 \ldots 0.081$ & $0.01 \ldots 0.03$ & $0.07 \ldots 0.1$ & $12$ \\
\mbox{i075} & $0.75$  & $399$  & $301$  & $0.081$ & $0.01 \ldots 0.03$ & $0.07 \ldots 0.1$ & $2$ \\
\mbox{i100} & $1.0$  & $27 \ldots 1582$ & $27   \ldots 1582$ & $0.069 \ldots 0.081$ & $0.01 \ldots 0.03$ & $0.07 \ldots 0.1$ & $14$ \\
\mbox{i200} & $2.0$  & $22$  & $44$  & $0.068$ & $0.01 \ldots 0.03$ & $0.07 \ldots 0.1$ & $2$ \\
\mbox{i500} & $5.0$  & $22.2$  & $111$  & $0.068$ & $0.01 \ldots 0.03$ & $0.07 \ldots 0.1$ & $2$ \\
\mbox{i1000} & $10.0$  & $22.3$  & $223$  & $0.068$ & $0.01 \ldots 0.03$ & $0.07 \ldots 0.1$ & $2$ \\
\mbox{du001}  & $0.01$ & $153$ & $ 1.5 $ & $ 0.078 $ & $ 0.01 $ & $ - $ & $10$ \\
\mbox{du010}  & $0.1$  & $153 \ldots 391$ & $15 \ldots 39 $ & $ 0.078 $ & $ 0.01 \ldots 0.03 $ & $ - $ & $11$ \\
\mbox{du100a} & $1.0$  & $154$ & $ 154 $ & $ 0.079 $ & $ 0.01 $ & $ - $ & $10$ \\
\mbox{du100b} & $1.0$  & $154$ & $ 154 $ & $ 0.079 $ & $ 0.03 $ & $ - $ & $10$ \\
\mbox{ds001}  & $0.01$ & $153$ & $ 1.5 $ & $ 0.078 $ & $ - $ & $ 0.1 $ & $10$ \\
\mbox{ds010}  & $0.1$  & $153 \ldots 391$ & $15 \ldots 39$ & $ 0.078 $ & $ - $ & $0.07 \ldots 0.1 $ & $11$ \\
\mbox{ds100}  & $1.0$  & $153$ & $ 153 $ & $ 0.078 $ & $ - $ & $ 0.1 $ & $10$ \\
  \end{tabular}
  \caption{Summary of runs. Runs with imposed velocity or specific
    entropy gradients are denoted with prefix i, whereas decay
    experiments of velocity (specific entropy) are identified by
    prefix du (ds). Grid resolutions range between $144^3$ and
    $1152^3$.}
  \label{tab:runs}
  \end{center}
\end{table}

\section{Results}
\label{sec:results}

The simulations discussed in the present study are listed in
\tab{tab:runs}.

\subsection{Turbulent viscosity and heat diffusion from imposed flow and entropy methods}
\label{imposed}

We measure the turbulent viscosity and thermal diffusivity in two
ways. First, we impose sinusoidal large-scale profiles of velocity
(\ref{equ:impu}) or entropy (\ref{equ:imps}). The response of the
system are non-zero Reynolds stress and vertical enthalpy flux
profiles that are parameterized with gradient diffusion terms
\citep[e.g.][]{R89}
\begin{eqnarray}
\Fenthz(z) = \cP \overline{(\rho u_z)' T'} \approx \cP \mean{\rho}
\overline{u_z' T'} = -\chi_{\rm t} \mean{\rho} \mean{T} \frac{\pd
  \mean{s}}{\pd z},
\end{eqnarray}
and
\begin{eqnarray}
\Ryz(z) = \overline{u_y'u_z'} = -\nu_{\rm t} \frac{\pd \mean{u}_y}{\pd z},
\end{eqnarray}
where primes denote fluctuations from the mean, e.g., $\uuu' = \uuu -
\mean{\uuu}$. The Mach number in the current simulations is always
less than 0.1. Therefore we neglect density-dependent terms in our
analysis because they scale with $\Ma^2$.

\begin{figure}
\begin{center}
\includegraphics[width=0.7\columnwidth]{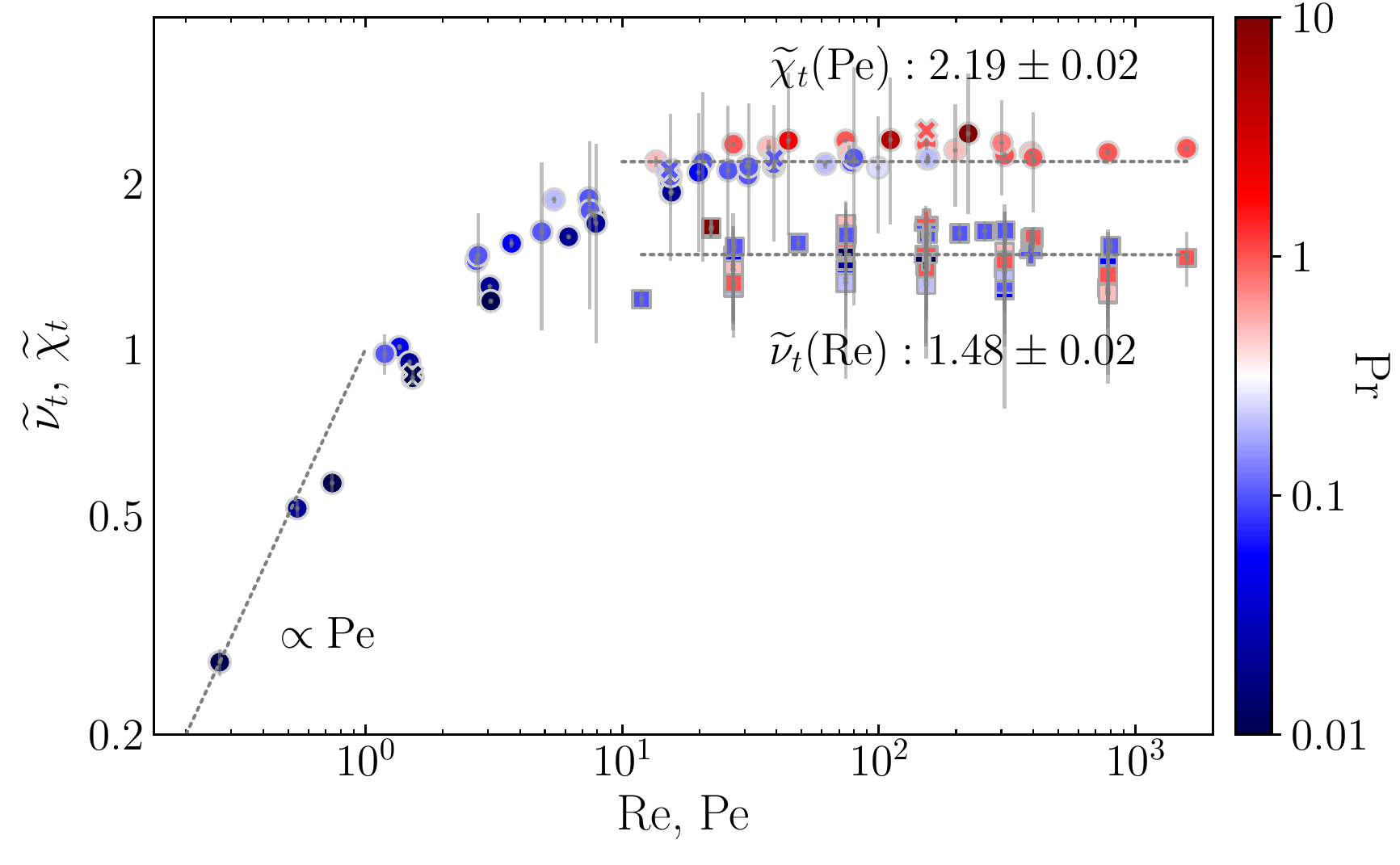}
\end{center}
\caption[]{Normalized turbulent viscosity $\tilde{\nut} = \nut/\nutz$
  (squares) and heat diffusivity $\tilde{\chit} = \chit/\chitz$
  (circles) as functions of Reynolds and P\'eclet numbers. The crosses
  ($\times$) and pluses ($+$) indicate results from decay
  experiments. The colours of the symbols indicate the microscopic
  Prandtl number as shown by the colourbar. The dotted horizontal
  lines show fit to the data for $\Pe,\Rey>10$ and a line proportional
  to $\Pe$ is shown for low $\Pe$.}\label{pr_prt1}
\end{figure}

The coefficients $\chit$ and $\nut$ are assumed to be scalars and
were obtained from linear fits
between time-averaged $\Fenthz$ and $\mean{\rho}\mean{T}\pd_z\mean{s}$
and between $\Ryz$ and $\pd_z \mean{u}_y$, respectively. Results from
our simulations are shown in \fig{pr_prt1}. We normalize $\nut$ and
$\chit$ by
\begin{eqnarray}
\nutz = \chitz = \onethird\urms\kf^{-1},
\end{eqnarray}
which is an order of magnitude estimate for the turbulent diffusion
coefficients. We note that in the parameter regimes studied here, the
estimates $\nutz$ and $\chitz$ are very similar in all of our runs.
Our results show that for low P\'eclet numbers the turbulent heat
diffusion increases in proportion to $\Pe$ for $\Pe \lesssim 1$. This
is consistent with earlier numerical results for turbulent viscosity
\citep[e.g.][]{2020A&A...636A..93K}, magnetic diffusivity
\citep[e.g.][]{SBS08}, and passive scalar diffusion \citep{BSV09}, and
with corresponding analytic results in the diffusion dominated
($\Pe\ll 1$) regime. For sufficiently large $\Pe$, $\tilde{\chit}$
tends to a constant value. The turbulent viscosity is also roughly
constant in the parameter space covered here. For low fluid Reynolds
numbers $\nut$ is proportional to $\Rey$ as has been shown in
\cite{2020A&A...636A..93K}. However, we do not cover this parameter
regime with the current simulations.

\subsection{Turbulent viscosity and heat diffusion from decay experiments}
\label{decay}

\begin{figure}
\includegraphics[width=0.5\columnwidth]{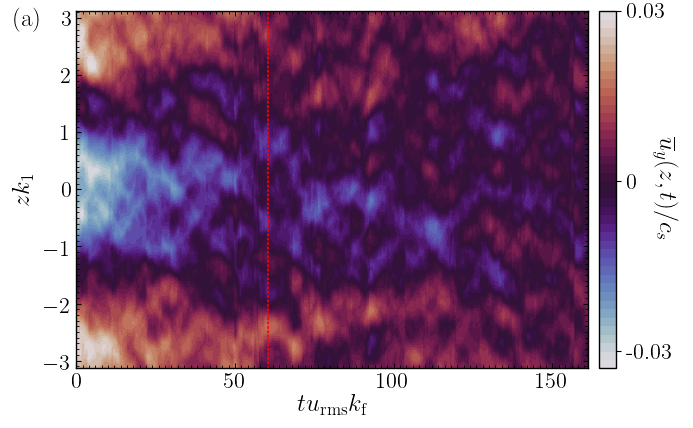}\includegraphics[width=0.5\columnwidth]{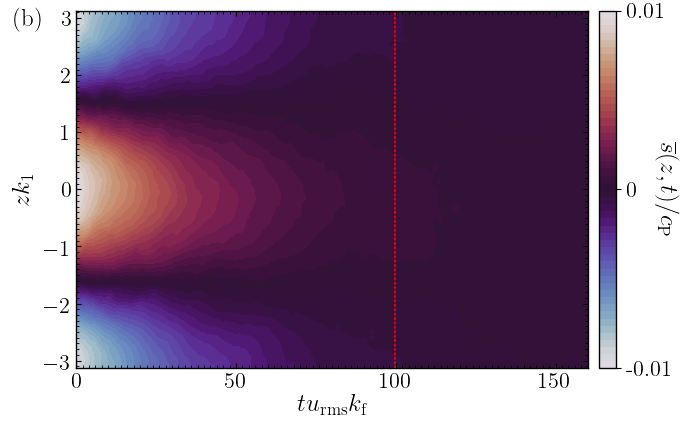}
\includegraphics[width=0.5\columnwidth]{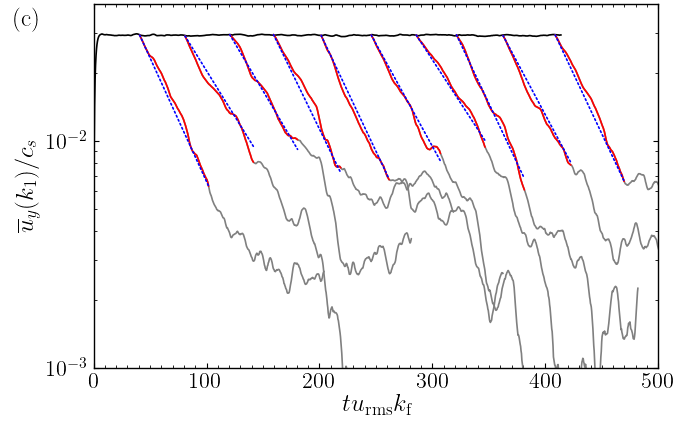}\includegraphics[width=0.5\columnwidth]{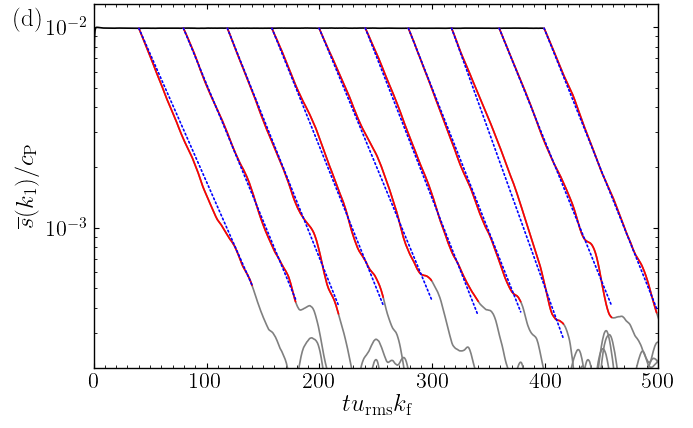}
\caption[]{Panels (a) and (b): $\mean{u}_y(t,z)$ and $\mean{s}(t,z)$
  normalized by $c_s$ and $\cP$, respectively, from decay experiments
  with $\Pra=1$ and $\Rey=157$. Red vertical lines denote end times of
  exponential fits. Panels (c) and (d): temporal decays of
  $k_z/k_1=1$ mode of $\mean{u}_y$ and $\mean{s}$, respectively; black
  line shows the progenitor run, and the red/gray lines indicate the
  decaying runs; red part is used to fit exponential decay; blue
  dotted lines show the exponential fit.}\label{puuss_decay_nut288_Pr1}
\end{figure}

Decay experiments were made as an independent way to measure the
turbulent viscosity and heat diffusion. Snapshots from the imposed
velocity/entropy gradient runs were used as initial conditions and the
relaxation terms of the rhs of the Navier--Stokes and
entropy equations were deactivated. The large-scale velocity and
entropy profiles in such runs decay due to the combined effect of
molecular and turbulent diffusion. To measure the decay rate we
monitored the amplitude of the $k = k_1$ components of $\mean{u}_y$
and $\mean{s}$. Exponential decay laws
\begin{eqnarray}
  \mean{u}_y(t,k_1) = \mean{u}_y(t_0,k_1) e^{-(\nut+\nu)t} ,\ \ \mean{s}(t,k_1) = \mean{s}(t_0,k_1) e^{-(\chit+\chi)t},
\end{eqnarray}
were then fitted to the numerical data. Representative examples from
decay experiments of large-scale velocity and entropy are shown in
\fig{puuss_decay_nut288_Pr1}. The upper panels (a) and (b) show
$\mean{u}_y(z,t)$ and $\mean{s}_y(z,t)$ from typical decay
experiments. The $k_1$ components of these fields decay exponentially
when the forcing is turned off; see panels (c) and (d) of
\fig{puuss_decay_nut288_Pr1}. Ultimately the amplitude of the $k_1$
mode decreases sufficiently such that it cannot be distinguished from
the background turbulence. The time it takes to reach this state
varies and depends on the initial amplitudes $u_0$ and $s_0$. However,
at the same time these amplitudes need to be kept as low as possible
to avoid non-linear effects becoming important \citep[see,
  e.g.][]{2020A&A...636A..93K}. This is particularly important for the
velocity field due to which the range from which turbulent viscosity
can be estimated is limited which necessitates running several
experiments with different snapshots as initial conditions to reach
converged values for $\nut$ and $\chit$.

Due to this, only a limited subset of the parameter range covered by
the imposed cases were repeated with decay experiments. We used ten
snapshots from each run for the decay experiments. The separation
between the snapshots is roughly $\Delta t = 40 \urms\kf$ such that
the realizations can be considered uncorrelated. Results from the
decay experiments are shown in \fig{pr_prt1} with crosses ($\chit$)
and pluses ($\nut$). We find that the results from the decay
experiments are consistent with those from the imposed flow and
entropy gradient methods.

\subsection{The $k-\epsilon$ model}

\begin{figure}
\begin{center}
\includegraphics[width=0.7\columnwidth]{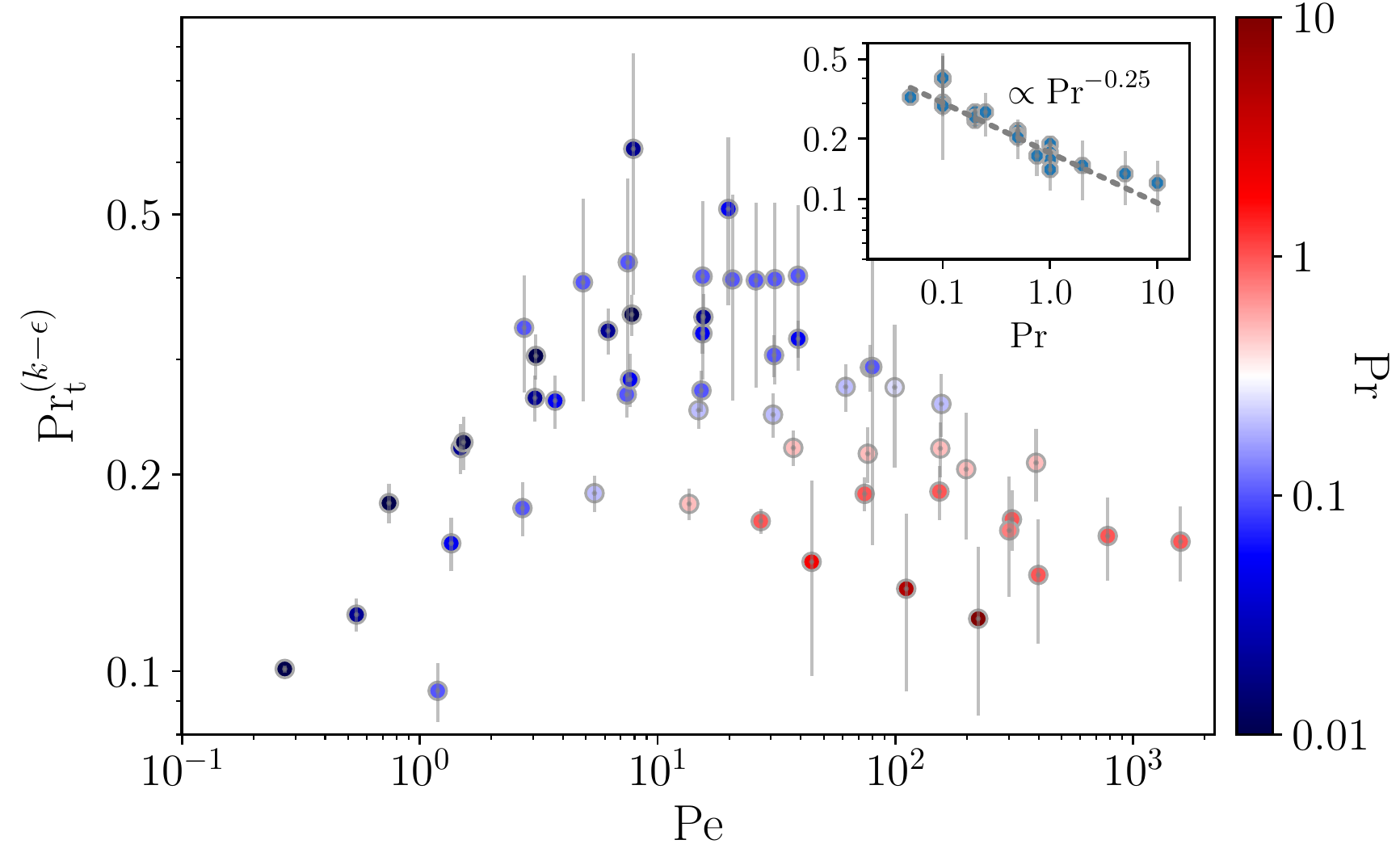}
\end{center}
\caption[]{Turbulent Prandtl number $\Pra_{\rm t}^{(k-\epsilon)}$
  according to (\ref{equ:Prt_ke}) as a function of P\'eclet
  number. The colour of the symbols denotes the molecular Prandtl
  number as indicated by the colour bar. Inset: $\Pra_{\rm
    t}^{(k-\epsilon)}$ versus $\Pra$ from runs with $\Pe>20$.}
\label{fig:pPrt_ke}
\end{figure}

To facilitate a comparison with \cite{2021PhRvF...6j0503P} we use the
expressions of $\nut$ and $\chit$ derived under the $k - \epsilon$
model with
\begin{eqnarray}
\nut^{(k-\epsilon)}  &=& c'_\nu k_u^2/\epsilon_K, \\
\chit^{(k-\epsilon)} &=& c'_\chi k_u k_T/\epsilon_T,
\end{eqnarray}
where $k_u = \brac{\uuu'^2}/2$ is the turbulent kinetic energy, $k_T =
\brac{T'^2}$ is the variance of the temperature fluctuations, $c'_\nu$
and $c'_\chi$ are assumed to be universal constants\footnote{Primes
  indicate the constancy of $c'_\nu$ and $c'_\chi$, but see
  \Sec{relax} where this assumption is lifted.}. Viscous and thermal
dissipation rates are defined as $\epsilon_K =
2\nu[\brac{\SSt^2}-\brac{\SSt_0}^2]$ and $\epsilon_T = \chi
\brac{(\bm\nabla T')^2} = \chi [\brac{(\bm\nabla T)^2} -
  \brac{(\bm\nabla\mean{T})^2}]$, respectively, where we have removed
contributions from the mean flow and the mean entropy; $\SSt_0$
denotes the traceless rate-of-strain tensor as defined in
(\ref{equ:SSt}) but with $\mean{\uuu}_0$ instead of $\uuu$.
\cite{2021PhRvF...6j0503P} computed $\Pra_{\rm t}$ using the
$k-\epsilon$ model by fixing the ratio of $c'_\nu/c'_\chi$, which
yields
\begin{eqnarray}
\Pra_{\rm t}^{(k-\epsilon)} = \frac{\nut^{(k-\epsilon)}}{\chit^{(k-\epsilon)}} =
\frac{c'_\nu}{c'_\chi} \frac{k_u \epsilon_T}{k_T \epsilon_u}.\label{equ:Prt_ke}
\end{eqnarray}
For simplicity, we assume $c'_\nu/c'_\chi=1$ in this subsection. The
results are shown in \fig{fig:pPrt_ke}. We find that taking the ratio
$c'_\nu/c'_\chi$ to be a constant leads to results where $\Pra_{\rm
  t}^{(k-\epsilon)}$ increases monotonically with decreasing $\Pra$
when $\Pe$ is larger than about 20; see the inset in \fig{fig:pPrt_ke}
which reveals a dependence of $\Pra^{-0.25}$.  Qualitatively, this
result is in agreement with the one in
\cite{2021PhRvF...6j0503P}. However, we would like to note here that a
strong assumption was made to reach this conclusion, namely, that the
ratio $c'_\nu/c'_\chi$ is fixed, and that $c'_\nu$ and $c'_\chi$ are
universal constants, independent of control parameters such as $\Rey$
and $\Pe$. Henceforth, we relax these assumptions, and also omit
primes from the coefficients $c_\nu$ and $c_\chi$.

\subsection{Relaxing the assumption that $c_\nu$ and $c_\chi$ are constants}
\label{relax}

\begin{figure}
\begin{center}
\includegraphics[width=0.7\columnwidth]{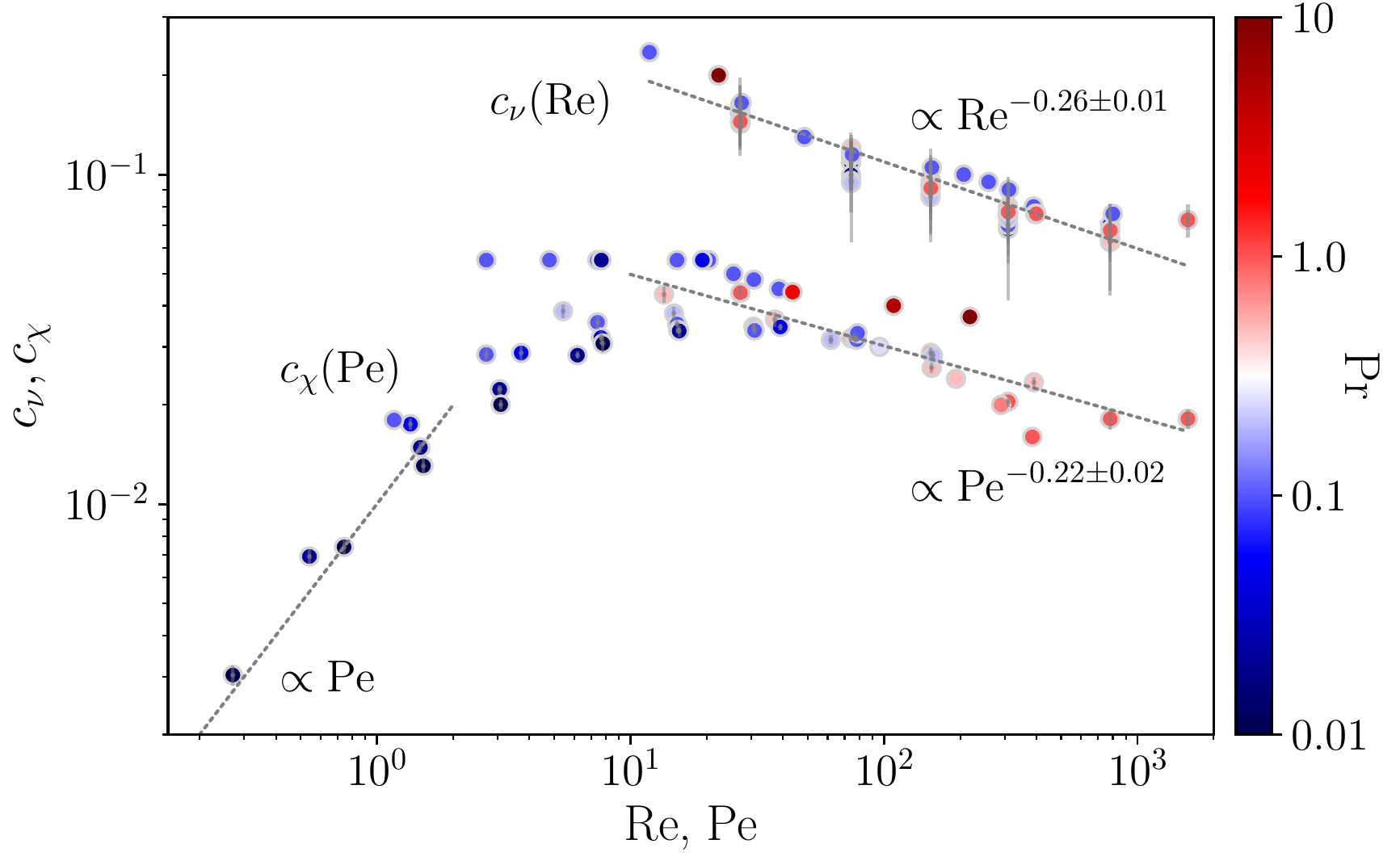}
\end{center}
\caption[]{Similar to \fig{pr_prt1} but for $c_\nu(\Rey)$ and
  $c_\kappa(\Pe)$ from (\ref{equ:cnu}) and (\ref{equ:cchi}). The
  colours again indicate the microscopic Prandtl number. The grey
  dotted lines indicate fits to the data for $(\Pe,\Rey>10)$ and a
  line proportional to $\Pe$ is shown for low $\Pe$.}\label{cnucchi}
\end{figure}

It is reasonable to assume that for sufficiently large Reynolds and
P\'eclet numbers $k_u$ and $k_T$ tend to constant values. Furthermore,
there is evidence from numerical simulations that $\epsilon_K$ also
tends to a non-zero constant value for large Reynolds numbers. Similar
evidence for $\epsilon_T$ has not been presented. Therefore it is not
clear whether the assumption of universality of $c_\nu$ and $c_\chi$
is valid. This is particularly important for numerical simulations
such as those in the current study where the Reynolds and P\'eclet
numbers are still modest. Since we have independently measured $\nut$
and $\chit$ using the imposed flow and entropy method (\Sec{imposed})
and from decay experiments (\Sec{decay}), we can estimate $c_\nu$ and
$c_\chi$ using:
\begin{eqnarray}
c_\nu &=& \nut/(k_u^2/\epsilon_K), \label{equ:cnu} \\
c_\chi &=& \chit/(k_u k_T/\epsilon_T), \label{equ:cchi}
\end{eqnarray}
where $\nut$ and $\chit$ are the ones obtained above in \Sec{imposed}
with the imposed field method. The results are shown in \fig{cnucchi}.
Our results indicate that $c_\nu$ and $c_\chi$ are highly variable and
that they depend not only on $\Rey$ and $\Pe$ but also on
$\Pra$. Furthermore, for sufficiently large Reynolds and P\'eclet
numbers, $c_\nu$ and $c_\chi$ show decreasing trends proportional to
roughly $-0.25$ power of $\Rey$ and $\Pe$, respectively. This shows
that any estimate of $\nut$ or $\chit$ with the $k-\epsilon$ model in
the parameter regime studied here would require prior knowledge of
$c_\nu$ and $c_\chi$ for the particular parameters ($\Rey$, $\Pe$,
$\Pra$) of that system.

\begin{figure}
\begin{center}
\includegraphics[width=0.7\columnwidth]{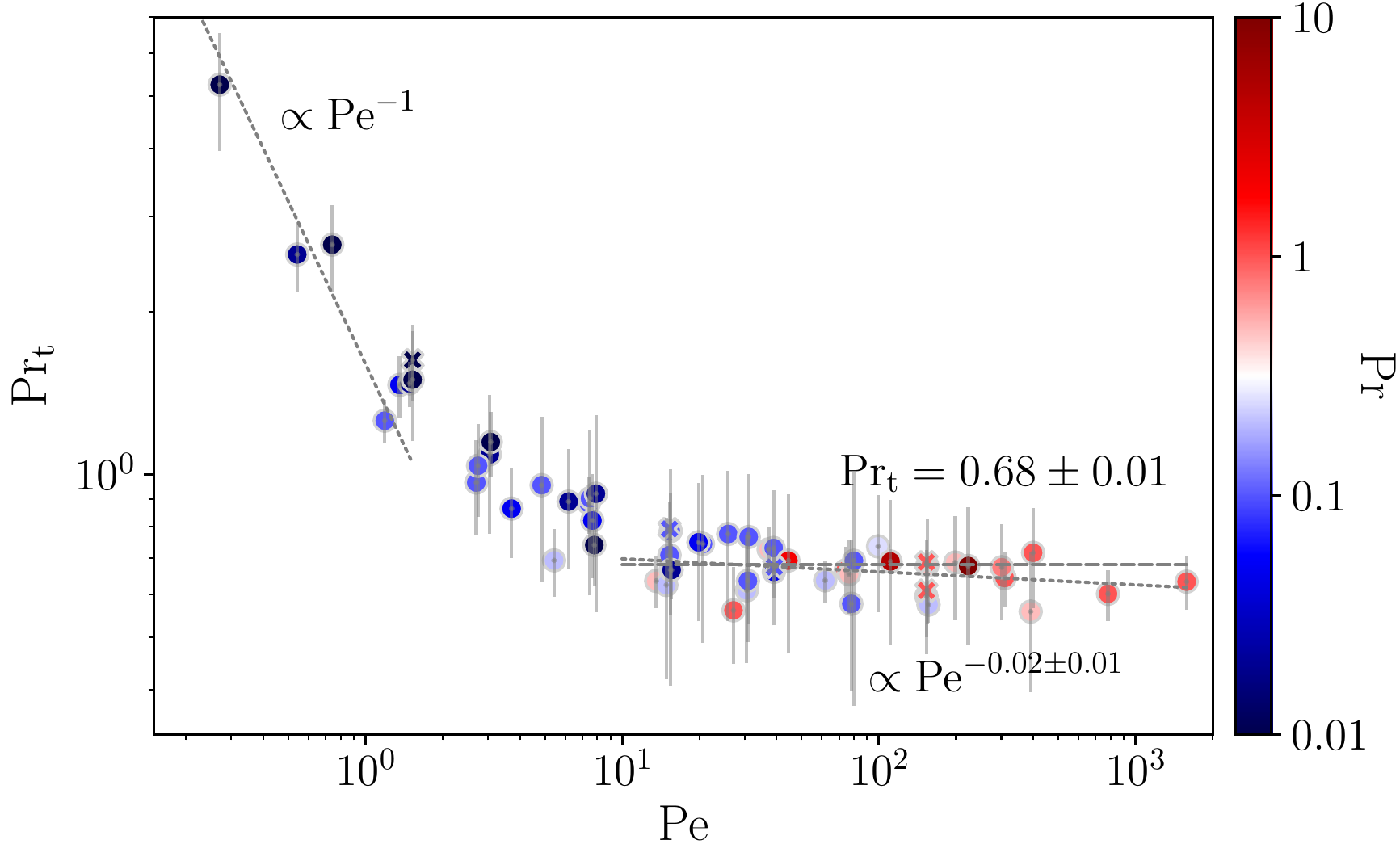}
\end{center}
\caption[]{Turbulent Prandtl number $\Pra_{\rm t} = \nut/\chit$ as a
  function of P\'eclet number. The colour of the symbols denotes the
  molecular Prandtl number as indicated by the colour bar. The crosses
  ($\times$) show results from decay experiments. Linear and power law
  fits to data for $\Pe>10$ are shown by the dashed and dotted lines,
  respectively, and a line proportional to $\Pe^{-1}$ is shown for low
  $\Pe$.}
\label{fig:pPrt}
\end{figure}

\subsection{Turbulent Prandtl number}

Our results for the turbulent Prandtl number
\begin{eqnarray}
\Pra_{\rm t} = \frac{\nut}{\chit}
\end{eqnarray}
are shown in \fig{fig:pPrt} with $\nu_t$ and $\chi_t$ as discussed in
\Secs{imposed}{decay}. We find that for $\Pe \lesssim 1$ the turbulent
Prandtl number is roughly inversely proportional to $\Pe$ for low
molecular Prandtl number. We have not computed the turbulent Prandtl
number for cases where both $\Rey$ and $\Pe$ are smaller than
unity. For sufficiently high P\'eclet number the turbulent Prandtl
number tends to a constant value which is close to 0.7. This is in
accordance with theoretical estimates in that $\Pra_{\rm t}$ is
somewhat smaller than unity. For example, \cite{R89} derived
$\Pra_{\rm t}=2/5$ using first-order smoothing approximation. The
turbulent Prandtl number plays an important role also in the
atmospheric boundary layer where several methods yield values of the
order of unity \citep[][and references therein]{LI201986}.

That, the turbulent Prandtl number $\Pra_{\rm t}$ reaches a constant
value at sufficiently large $\Pe$, independent of $\Pra$, is in stark
contrast to the results obtained from the $k-\epsilon$ model with a
fixed $c_\nu/c_\chi$; compare \figs{fig:pPrt_ke}{fig:pPrt}. Now we
make an attempt to understand the reason for this discrepancy.  From
\fig{cnucchi} we note the following approximate scaling relations at
sufficiently large $\Rey$ and $\Pe$: $c_\nu \propto \Rey^{-0.25}$ and
$c_\chi \propto \Pe^{-0.25}$, suggesting thus that the ratio
$c_\nu/c_\chi$ scales with the Prandtl number as $\Pra^{+0.25}$.  With
this, if we let $c'_\nu/c'_\chi \propto \Pra^{+0.25}$ in
\ref{equ:Prt_ke}, instead of a fixed ratio, and note from the inset of
\fig{fig:pPrt_ke} that the factor $k_u \epsilon_T/k_T \epsilon_u
\propto \Pra^{-0.25}$, we would obtain from \ref{equ:Prt_ke} that
$\Pra_{\rm t}^{(k-\epsilon)}$ becomes independent of $\Pra$, agreeing
thus qualitatively with our results as shown in
\fig{fig:pPrt}. Therefore we conclude that the results from the
$k-\epsilon$ model with a fixed value for the ratio $c_\nu/c_\chi$ are
unreliable and that the strong dependence of $\Pra_{\rm t}$ on $\Pra$
found in \cite{2021PhRvF...6j0503P} is due to the restrictive
assumption in their model.

\section{Conclusions}
\label{sec:conclusions}

Using simulations of weakly compressible isotropically forced
turbulence with imposed large-scale gradients of velocity and
temperature, and corresponding decay experiments, we find that the
turbulent Prandtl number $\Pra_{\rm t}$
is roughly 0.7 and independent of the microscopic Prandtl number
$\Pra$ provided that the Pecl\'et number is higher than about
ten. This is in stark contrast from the recent results of
\cite{2021PhRvF...6j0503P} who found that $\Pra_{\rm t} \propto
\Pra^{-1}$ from non-Boussinesq simulations of convection. Although the
physical setups are quite different, we were able to qualitatively
reproduce their finding under the strong assumption that
$c_\nu/c_\chi$ is fixed, and note that the method by which the
turbulent viscosity and thermal diffusivity were obtained in
\cite{2021PhRvF...6j0503P} produce unreliable results even in the
simpler cases considered here. Relaxing their assumption of
universality of $c_\nu$ and $c_\chi$, we find that these
depend not only on $\Rey$ and $\Pe$, respectively, but also on $\Pra$.
This allows us to understand the reason for the discrepancy.

\backsection[Acknowledgments]{We thank S. Sridhar for his comments on
  an earlier draft of the manuscript. NS thanks the hospitality
  provided by RRI, Bangalore where parts of this work were done. We
  gratefully acknowledge the use of high performance computing
  facilities at HLRN in G\"ottingen and Berlin, and IUCAA, Pune.}

\backsection[Funding]{This work was supported by the Deutsche
  Forschungsgemeinschaft Heisenberg programme (P.J.K., grant number
  KA4825/4-1).}

\backsection[Declaration of interests]{The authors report no conflict
  of interest.}

\backsection[Author ORCID]{P. J. K\"apyl\"a,
  https://orcid.org/0000-0001-9619-0053; N. K. Singh,
  https://orcid.org/0000-0001-6097-688X.}

\backsection[Author contributions]{Both authors contributed equally to
  running the simulations, analyzing data, reaching conclusions, and
  in writing the paper.}

\bibliographystyle{jfm}
\bibliography{paper}

\end{document}